# Structure and physical properties of the noncentrosymmetric superconductor $Mo_3Al_2C$

A.B. Karki, Y.M. Xiong, I.Vekhter, D. Browne, P.W. Adams, and D.P. Young

*Department of Physics and Astronomy, Louisiana State University, Baton Rouge, Louisiana, 70803, USA*

K. R. Thomas and Julia Y. Chan

*Department of Chemistry, Louisiana State University, Baton Rouge, Louisiana, 70803, USA*

R. Prozorov and H. Kim

*Department of Physics and Astronomy, Iowa State University, Ames, Iowa, 50011, USA*

We have synthesized polycrystalline samples of the noncentrosymmetric superconductor $Mo_3Al_2C$ by arc and RF melting, measured its transport, magnetic and thermodynamic properties, and computed its band structure. Experimental results indicate a bulk superconducting transition at $T_c \sim 9.2$ K, while the density of states at the Fermi surface is found to be dominated by Mo $d$-orbitals. Using the measured values for the lower critical field $H_{c1}$, upper critical field $H_{c2}$, and the specific heat $C$, we estimated the thermodynamic critical field $H_c(0)$, coherence length $\xi(0)$, penetration depth $\lambda(0)$, and the Ginzburg-Landau parameter $\kappa(0)$. The specific heat jump at $T_c$, $\Delta C/\gamma T_c = 2.14$, suggests that $Mo_3Al_2C$ is moderately-to-strongly coupled, consistent with the fast opening of the gap, as evidenced by the rapid release of entropy below $T_c$ from our electronic specific heat measurements. Above 2K the electronic specific heat exhibits the power law behavior, suggesting that synthesis of single crystals and measurements at lower temperature are needed to establish whether the gap is anisotropic. The estimated value of the upper critical field $H_{c2}(0)$ is close to the calculated Pauli limit, therefore further studies are needed to determine whether the absence of an inversion center results in a significant admixture of the triplet component of the order parameter.

## I. INTRODUCTION

Metal carbide and nitride intermetallic superconductors have both fairly high transition temperatures and large critical magnetic field values. Some of the metal carbides, for example $MgCNi_3$, $La_2C_3$, $Y_2C_3$, $Mo_3C_2$, $LaNiC_2$, $Mo_3Al_2C$ etc., exhibit exotic physical, structural and electronic properties [1-10]. Furthermore, $La_2C_3$, $Y_2C_3$, $LaNi_2C$, and $Mo_3Al_2C$ have an additional facet of complexity, in that their crystal structures lack a center of inversion symmetry [3,4,7,9]. In such systems the antisymmetric (in the electron momentum) spin-orbit coupling breaks parity, so that spin is no longer a good quantum number. Instead, the quasiparticle states are classified by their helicity, the relative orientation of the spin and crystal momentum. Consequently, the Fermi surface is split into sheets of different helicity by the spin-orbit interaction. Superconducting pairing on the same Fermi surface sheet (the states of the same helicity) is therefore a mixture of spin singlet and triplet, and may exhibit unusual magnetic properties and/or have nodes in the energy gap, depending on the structure and the strength of the triplet component[11-15]. Importantly, in contrast to the triplet *p*-wave superconductors proposed in centrosymmetric systems, such as $Sr_2RuO_4$, lack of inversion symmetry makes it possible for the triplet component to exist in a spatially fully symmetric paired state, so that one can speak, for example, of triplet *s*-wave superconductor [12]. Similarly, nodes in the superconducting gap may exist even when the pair wave function exhibits the full spatial symmetry of the crystal [12-16].

Studies of pairing in such systems have been triggered by the discovery of superconductivity in the heavy-fermion $CePt_3Si$ [17,18]. The subsequent investigations into the pairing symmetry of this class of superconductors [12,14,19,20] included $CeIrSi_3$, UIr, $Li_2(Pd Pt)_xB$, and $Re_3W$ [20-26]. Some of these systems, such as $Li_2Pd_3B$, behave very much like conventional fully-gapped singlet superconductors, indicating that the triplet admixture is weak. On the other hand, the isostructural $Li_2Pt_3B$ shows evidence for line nodes in its energy gap [16].

In this paper we focus on the noncentrosymmetric superconducting carbide $Mo_3Al_2C$. To our knowledge, there have been no comprehensive reports on its electronic structure and physical properties at low temperature. Johnston *et al*. reported the synthesis of $T_3Al_2C$ ($T$ = Mo, Nb, Ta, V, Ti, Cr) employing various techniques and determined the crystal structure and superconducting properties above 4.2 K [10]. All the compounds have the cubic *β*-Mn (A13) structure, but they found that only $Mo_3Al_2C$ exhibits superconductivity at a moderately high temperature near 10 K. This compound was synthesized by hot pressing the metal powders in graphite molds. The reported lattice constant was 6.867 Å. Later, Fink *et al.* reported the upper critical field ($H_{c2}$) = 15.6 T at 1.2 K, which was determined from critical current density measurements [27]. From the experimentally determined values of the upper critical field ($H_{c2}$), the normal-state resistivity ($\rho_n$), the critical current density ($J_c$), and the transition temperature ($T_c$), they estimated the electronic specific heat coefficient ($\gamma$), lower critical field ($H_{c1}$), and Ginzburg-Landau parameter ($\kappa$) at 1.2 K [27]. Several carbides and nitrides with isotypic H-phases and phases with the filled and ordered *β*-Mn structure were tested for superconductivity by L.E. Toth *et al.* [28]. Only the *β*-Mn compound $Nb_3Al_2N$ was found to be superconducting at 1.3 K, confirming a previous result that the filled *β*-Mn crystal structure is favorable for superconductivity.

Here we report on the synthesis and characterization of polycrystalline $Mo_3Al_2C$ which shows a bulk superconducting transition at $T_c \sim 9.2$ K. Measurements of the transport, magnetic, and thermal properties, as well as a calculation of its electronic structure, have been performed. From these measurements, we estimate various superconducting parameters of $Mo_3Al_2C$, including the thermodynamic critical field $H_c(0)$, coherence length $\xi(0)$, penetration depth $\lambda(0)$, electronic specific heat coefficient $(\gamma)$, and the specific heat jump $\Delta C/T_c$.

## II. EXPERIMENT

A polycrystalline sample (0.3 g) of $Mo_3Al_2C$ was prepared by melting stoichiometric amounts of high purity molybdenum powder (99.95 % Alfa Aesar), aluminum shot (99.999 % Alfa Aesar), and carbon graphite rod (99.999 % Alfa Aesar) under flowing ultra high purity argon gas in an arc melter using a tungsten electrode and a water-cooled copper hearth. A Zr button was used as an oxygen getter. First, a button of Mo was made by arc melting the Mo powder alone. The Mo button was then melted with Al pieces with 20 % excess to compensate for the loss of some Al due to its higher vapor pressure. Finally, the arc melted button of Mo-Al was re-melted with pieces of carbon graphite rod. The sample was re-melted several times to ensure homogeneous mixing of the constituent elements. Another sample was made by melting a cold-pressed pellet of a well-ground mixture of Mo, Al and C powders in a stoichiometric ratio in an RF-induction furnace. The sample formed a hard pellet with a negligible mass loss. We made more than 10 samples by various methods and with different concentrations of C and Al in order to synthesize the stoichiometric compound and to check the effects of off-stoichiometry on $T_c$.

X-ray diffraction analysis on well-ground powder of a portion of the sample was carried out on a Bruker Advance D8 powder diffractometer equipped with Cu $K_\alpha$ radiation ($\lambda$ = 1.54056 Å). Data were collected from $2\theta = 10^o$ to $70^o$ with a constant scan speed of $2^o$ min$^{-1}$ at room temperature.

The electrical resistivity was measured using a standard 4-probe ac technique at 27 Hz with an excitation current of 3 mA, in which small diameter Pt wires were attached to the sample using a conductive epoxy (Epotek H20E). Data were collected from 1.8 K to 290 K and in magnetic fields up to 9 T using a Quantum Design, Physical Property Measurement System (PPMS). The specific heat was measured in the PPMS using a time-relaxation method between 2 K and 20 K at 0 T and 9 T. Magnetic susceptibility was measured in a constant magnetic field of 30 Oe; the sample was cooled down to 3 K in zero field, and then magnetic field was applied, followed by heating to 10 K (ZFC) and then cooled down again to 3 K (FC) in the PPMS.

## III. RESULTS AND DISCUSSIONS

The XRD pattern of the RF-melted sample of $Mo_3Al_2C$ is shown in Figure 1. The pattern indicates the sample is nearly single phase, with a small amount of $Mo_3Al_8$ and elemental Mo present as impurities. By least-square fitting to the positions of 7 Bragg reflection peaks between $2\theta$ values of 10° and 70°, the cubic cell parameter a = 6.866 Å was obtained, consistent with the previous report [10]. A schematic view of the crystal structure of $Mo_3Al_2C$ is presented in the inset of Figure 1. It assumes a filled *β*-Mn structure, ideally consisting of a lattice of corner-sharing $Mo_6C$ octahedra in a five-member ring arrangement and an interpenetrating net-like lattice of Al atoms which occupy the pentagonal holes formed by the molybdenum carbide polyhedral rings. The primitive cubic Bravais lattice adopts the space group $P4_132$ and lacks a center of inversion symmetry.

The band structure of $Mo_3Al_2C$ was calculated using the WIEN2K full potential LAPW band package, [29] using the GGA exchange correlation potential [30]. The muffin tin radii were taken as 2.13 au for Mo, 2.42 au for Al, and 1.89 au for C. The cutoff in the LAPW basis was varied from $R*K_{max} = 6$ (1196 plane waves) to $R*K_{max} = 8$ (2795 plane waves). The mesh of *k*-points in the irreducible wedge of the Brillouin zone included 286 points, giving a total of 10,000 points in the full Brillouin zone. Calculations were performed using the room temperature lattice constant of 6.860 Å. The results of these calculations are presented in Figure 2 for the bands and Figure 3 for the density of states. The states at the Fermi level are derived almost entirely from the *d* bands of molybdenum. The initial iterative solution of the DFT equations omitted the spin-orbit interaction. The spin-orbit interaction was then added in the scalar relativistic approximation and the solution iterated again to convergence. Because of the relative small mass of the constituents, the spin orbit leads only to a splitting of the d-level degeneracies at high symmetry points. The splitting amounts to 0.02 eV at most. However, since none of those degeneracies occur close to the Fermi level, there are only small changes in the Fermi surface and the density of states near the Fermi level.

The temperature dependence of the normal state resistivity of $Mo_3Al_2C$ between 10 K and 290 K is shown in Figure 4. The resistivity is metallic, and one observes an inflection point in $\rho(T)$ at ~150 K. The normal-state resistivity is fairly low (~190 $\mu\Omega$ cm at room temperature), although it is of polycrystalline form. However, the residual resistivity ratio ($\rho_{290 K} / \rho_{10 K}$ = 1.09) is small, suggesting either the resistivity has little temperature dependence, or the transport in the polycrystalline sample is sensitive to impurities and disorder. Subsequent annealing of earlier samples actually worsened their crystalline quality, suggesting that the $Mo_3Al_2C$ phase is quenched in at high temperature. We have fit the low temperature resistivity data to a power law,

$$\rho = \rho_0 + aT^2, \qquad (1)$$

with the residual resistivity $\rho_0 = 177.35$ $\mu\Omega$ cm and coefficient $a = 0.38$ $\mu\Omega$ cm / $K^2$. The fit is shown as a solid line in Figure 4, and describes the data well between 10 K and 114 K. The inset of the figure shows the linear relationship between resistivity and temperature squared over the above range of temperature. A semi-log plot of the resistivity of polycrystalline $Mo_3Al_2C$ over the entire range of temperature from 3 K to 290 K is presented in Figure 5. The inset of Figure 5 shows the resistivity at lower temperature. The resistive superconducting transition is very sharp. The midpoint of the transition is 9.1 K, and the 90% −10% transition width is less than 0.2 K, with an onset temperature of 9.2 K.

The magnetic characterization of the superconducting transition at 30 Oe is shown in upper panel of Figure 6 a. A large diamagnetic superconducting signal is observed near 9.1 K, slightly lower than that in the resistivity. The superconducting volume fraction of the arc-melted spherical button was ~ 100 %, after applying the demagnetization factor to the ZFC data at 3 K. As shown in the figure, the sample has reached a full Meissner state at a temperature as high as 6 K. The magnetization vs. temperature data over the range 10 K to 300 K at 1000 Oe (not shown here) shows temperature-independent Pauli-paramagnetic behavior with a very small moment.

The magnetization vs. magnetic field over a range of temperatures below $T_c$ is presented in the Figure 6 b. For the analysis, the minimum point of the data curve was chosen as the lower critical field at each temperature, and the resulting $\mu_0H_{c1}(T)$ is plotted in Figure 7. A least squares fit to the data using

$$H_{c1}(T) = H_{c1}(0)[1-(T/T_c)^2] . \qquad (2)$$

yielded $\mu_0H_{c1}(0) = 0.0047$ T.

The magnetic field-dependent resistivity of $Mo_3Al_2C$ was measured for the same sample between 2 K and 10 K. The data are shown in the inset of Figure 8. As the field is increased, the superconducting transition becomes broader, and $T_c$ shifts to lower temperature. The mid-point of the resistivity transition was chosen as the transition temperature at each applied magnetic field. It should be noted that a sharp transition to a zero resistance state was obtained even at 9 T and above 4.5 K, indicating a high upper critical field. In the main panel of Figure 8, the upper critical field is plotted as a function of temperature. The variation in $H_{c2}$ with temperature is nearly linear with a negative slope, and it does not show any kind of saturation for fields as high as 9 T. We estimate $\mu_0H_{c2}(0)$ from the Werthamer-Helfand-Hohenberg (WHH)[31] expression,

$\mu_0 H_{c2}(T) = -0.693\, \mu_0 (dH_{c2}/dT)_{T=T_c}\, T_c$ . (3)

From the estimate of the slope, $\mu_0 dH_{c2}/dT \approx -2.39$ T / K and using $T_c = 9.2$ K, we found $\mu_0 H_{c2}(0) = 15.08$ T. This is in qualitative agreement with the upper critical field of the $Mo_3Al_2C$ reported by Fink et al. to be 15.6 T at 1.2 K [27]. A complementary estimate of the upper critical field from the empirical formula

$H_{c2}(T) = H_{c2}(0)(1-t^2)/(1+t^2)$, (4)

where $t = T/T_c$, gives $\mu_0 H_{c2}(0) = 18.2$ T, as shown in Figure 8 by the solid line representing the best fit to the experimental data. This extrapolated value is slightly larger than that obtained from the WHH theory.

The characterization of the superconducting transition by specific heat measurements is shown in Figure 9. The main panel shows the variation of $C/T$ with $T^2$ at lower temperature at a magnetic field of 0 and 9 T. The superconducting transition is suppressed to 5.2 K with the application of a 9-T magnetic field. This is consistent with the result from the resistivity data. The data above 5.5 K provide an extrapolation of the normal-state behavior to the $T = 0$ limit, and allow the determination of the Sommerfeld constant ($\gamma$) from the fit $C(T)/T = \gamma + \beta T^2$, where $\beta T^3$ is the lattice contribution to the specific heat. The extrapolation produced $\gamma = 18.65$ mJ mol$^{-1}$K$^{-2}$ and $\beta = 0.3050$ mJ mol$^{-1}$K$^{-4}$. Taking $\beta = N(12/5)\pi^4 R \Theta_d^{-3}$, where $R = 8.314$ J mol$^{-1}$ K$^{-1}$, and $N = 6$ for $Mo_3Al_2C$ gives a Debye temperature of $\Theta_d = 338.52$ K.

In the inset of Figure 9, $C/T$ is plotted as a function of $T$ showing the thermodynamic transition, as well as the jump $\Delta C/T_c$. The midpoint of the superconducting transition temperature is ~9.1 K, which is close to the value from the resistivity and magnetic measurements. It is known that the ratio $\Delta C/\gamma T_c$ can be used to measure the strength of the electron coupling. The specific heat jump ($\Delta C/T_c$) for $Mo_3Al_2C$ is 40 mJ mol$^{-1}$K$^{-2}$, which sets the value of $\Delta C/\gamma T_c$ to 2.14. This is higher than the BCS value of 1.43 for a weakly coupled superconductor, suggesting that $Mo_3Al_2C$ may be strongly-coupled. This is supported by the rapid decrease of the electronic contribution to the specific heat below the transition, signifying a substantial release of entropy in the immediate vicinity of the transition (Fig. 10 a).

Based on the specific heat data and the extrapolated value of the upper critical field $\mu_0 H_{c2}(0)$ obtained above, we can estimate the superconducting parameters of $Mo_3Al_2C$. The condensation energy, $U(0)$, was estimated from [32]

$$\Delta U(0) = -\frac{\gamma T_c^2}{2} + \int_0^{T_c} C_{es}(T) dT , \quad (5)$$

subject to the entropy-conserving fit for the specific heat.

Using the fit equation $C_{es} = a + bT^3$, we obtained $U(0)$ to be 465.92 mJ/mole. Then from $\gamma = 2\pi^2/3\, k_B^2\, N(0)$ and $U(0) = \frac{1}{2} N(0) \Delta^2(0)$, where $N(0)$ is density of states at the Fermi surface, and $2\Delta(0)$ is the superconducting gap energy, we estimated $\Delta(0)$ to be $25 \times 10^{-23}$ J, or $2\Delta_0 = 4.028 k_B T_c$. The enhanced value of this ratio agrees with strongly-coupled nature of the superconductivity as suggested by the specific heat jump.

Assuming the $g$-factor for the conduction electrons is close to 2, we estimate the Pauli limiting field for $Mo_3Al_2C$ from $\mu_0 H^{Pauli} = \Delta_0/\mu_B\sqrt{2}$, to be 19 T. This value of $\mu_0 H^{Pauli}$ is close to the estimated orbital field $\mu_0 H_{c2}(0) \approx 15\text{-}18$T. It is therefore highly desirable to extend the investigations of the upper critical field to lower temperatures to determine its detailed behavior. If the upper critical field exceeds the Pauli limit, it would signify a substantial contribution from the triplet component to the pairing amplitude, which would most likely be due to the broken inversion symmetry. Further studies of the anisotropy of the critical field in single crystals would be even more enlightening, since they would shed light on the structure of the spin-orbit coupling in the momentum space [13].

If we assume the upper critical field to be purely orbital, the superconducting coherence length can be estimated using $H_{c2}(0) = \Phi_0/2\pi\xi(0)^2$, where $\Phi_0 = 2.0678 \times 10^9$ Oe Å$^2$ is a flux quantum [33], to give $\xi(0) = 42.53$ Å and 46.84 Å for $H_{c2}(0) = 18.2$ T and 15.08 T, respectively. The value for the coherence length in $Mo_3Al_2C$ is very close to that for $MgCNi_3$ (46 Å) and is shorter than the coherence length in $Mo_3C_2$ (76 Å), [34,8]. Similarly, from the relation $H_{c1}(0) = (\Phi_0/4\pi\lambda^2)ln(\lambda/\xi)$, we find the magnetic penetration depth $\lambda(0) = 3755$ Å. The Ginzburg-Landau parameter is then $\kappa = \lambda/\xi = 88.56$. The thermodynamical critical field $H_c(0)$ was also estimated through the relation $U(0) = (½)\mu_0 H_c(0)^2$. The estimation resulted in $\mu_0 H_c(0) = 0.14$ T. Our experimental values of $H_{c2}(0)$, $H_{c1}(0)$, and $H_c(0)$ satisfy the relation $H_{c2}(0) H_{c1}(0) = H_c(0)^2[\ln\kappa(0) + 0.08]$ for $\kappa \gg 1$. [35]

The experimentally determined values of all parameters of our sample and other reference compounds are listed in the Table 1.

The behavior of the linear coefficient in the electronic specific heat at low temperatures, $T \ll T_c$, is an indicator of the superconducting gap structure. Power laws in $C_{es}/T$ generally indicate the existence of gapless excitations and nodes in the gap, while the exponential behavior, $C_{es}/T \approx a\ \exp(-\Delta/k_BT)$ indicates a fully gapped material. Presence of a triplet component of pairing, due to the absence of inversion center, would generally lead to an anisotropic gap. In determining the condensation energy above, we used a power law fit to the electronic specific heat, $C_{es}$, of $Mo_3Al_2C$ in the superconducting state down to 2.5 K (Figure 10 a). However, this is the high temperature regime that includes the peak at the transition, and this power law only indicates a rapid opening of the gap below the transition, without providing information on the nodal structure. For T < 2.5 K, $C_{es}$ varies nearly exponentially with T as shown in Figure 10 c, with the activation energy close to the gap value inferred from our analysis, suggesting a full gap. However, our sample is polycrystalline. In addition, an anomaly around 2.5 K, which corresponds to the superconducting transition temperature of $Mo_2C$, was detected by the penetration depth measurements, even though it is not seen in the specific heat data on the same sample, suggesting that the volume fraction of the impurity phase is small. In some of the other samples of $Mo_3Al_2C$, the presence of $Mo_2C$ had been detected by XRD, resistivity, magnetization and specific heat measurements too. The $C_{es}$ of pure $Mo_2C$ follows the BCS exponential relation below $T_c$. Therefore, while the data below 2.5 K are consistent with a full energy gap, we note that impurity scattering may have resulted in the gap being more isotropic than it would be in a pure sample. Measurements on pure single crystals of $Mo_3Al_2C$ are needed to definitively determine the gap structure.

## IV. CONCLUSION

We prepared polycrystalline samples of $Mo_3Al_2C$ by RF-induction and arc-melting techniques. The sample melted by induction was nearly single phase. The structure analysis shows that it lacks inversion symmetry. From resistivity, magnetic susceptibility, and specific heat measurements $Mo_3Al_2C$ was confirmed as a type II superconductor with a transition temperature near 9.2 K. Since the upper critical field $H_{c2}(0)$ is close to the estimated Pauli limiting field, further investigations into the possible mixture of single and triplet pairing states need to be done. The high value of $\Delta C/\gamma T_c \sim 2.14$ indicates that the compound is in moderate coupling regime. The electronic specific heat shows that a large portion of the entropy is released just below $T_c$, and the gap opens quickly to the value $\Delta_0/k_B T_c = 4.028$. While this supports classifying $Mo_3Al_2C$ as a reasonably strongly-coupled superconductor, it could also indicate the existence of an anisotropic gap. In the range of 2 K to 8 K, the electronic specific heat, $C_{es}$, follows the power law which, while consistent with rapid gap opening below the transition, would indicate an unconventional gap if found to extend to lower temperatures in single crystal samples. As it is, at low temperature our best fit is exponential, but the impurity effects may have masked the underlying gap anisotropy. The availability of single crystals is highly desirable to help answer these questions.

## V. ACKNOWLEDGMENTS

P.W.A. acknowledges the support of DOE under Grant No. DE-FG02-07ER46420, D.P.Y. and J.Y.C acknowledge the support of the NSF under Grant No. DMR-0449022 and DMR-0756281, respectively, and I. Vekhter acknowledges the support of the DOE under Grant No. DE-FG02-08ER46492.

[1] T. He, Q. Huang, A. P. Ramirez, Y. Wang, K. A. Regan, N. Rogado, M. A. Hayward, M. K. Haas, J. S. Slusky, K. Inumara, H. W. Zandbergen, N. P. Ong, and R. J. Cava, Nature ~London **411**, 54 (2001)

[2] S. Y. Li, R. Fan, X. H. Chen, C. H. Wang, W. Q. Mo, K. Q. Ruan, Y. M. Xiong, X. G. Luo, H. T. Zhang, L. Li, Z. Sun, and L. Z. Cao, Physical Review B **64**, 132505 (2001)

[3] K. Sugawara, T. Sato, S. Souma, T. Takahashi, and A. Ochiai, Physical Review **76**, 132512 (2007)

[4] G. Amano, S. Akutagawa, T. Muranaka, Y. Zenitani, and J. Akimitsu, J. Phys. Soc. Jpn. **73**, 530 (2004)

[5] A. Simon, and T. Gulden, Z. Anorg. Allg. Chem. **630**, 2191 (2004)

[6] A. Harada, S. Akutagawa, Y. Miyamichi, H. Mukuda, Y. Kitaoka, and J. Akimitsu, J. Phys. Soc. Jpn. **76**, 023704 (2007)

[7] I. R. Shein and A. L. Ivanovskii, Solid State Commun. **131**, 223 (2004)

[8] K.Yamaura, Q. Huang, M. Akaishi, and E. Takayama-Muromachi, Physical Review B **74**, 184510 (2006)

[9] W.H. Lee, H.K. Zeng, Y.D. Yao, Y. Y. Chen, Physica C **266**, 138-142 (1996)

[10] J. Johnston, L. E. Toth, K. Kennedy, and E. R. Parker, Solid State Commun. **2**, 123 (1964)

[11] L. P. Gorkov, and E.I Rashba, Phys. Rev. Lett. **87**, 037004 (2001)

[12] P.A. Frigeri, D.F.Agterberg, A. Koga, M. Sigrist, Phys. Rev. Lett. **92**, 097001 (2004)

[13] Manfred Sigrist, D. F. Agterberg, P. A. Frigeri, N. Hayashi, R. P. Kaur, A. Koga, I. Milat, K. Wakabayashi, and Y. Yanase, Journal of Magnetism and Magnetic Materials **310**, 536 (2007)

[14] P.A. Frigeri, D. F. Agterberg, I. Milat, and M. Sigrist, Eur. Phys. Journal **B 54**, 435 (2006)

[15] P. A. Frigeri, D. F. Agterberg, and M. Sigrist, New J. Phys. **6**, 115 (2004)

[16] H.Q. Yuan, D. F. Agterberg, N. Hayashi, P. Badica, D. Vandervelde, K. Togano, M. Sigrist, and M. B. Salamon, Phys. Rev. Lett. **97**, 017006 (2006)

[17] E. Bauer, G. Hilscher, H. Michor, Ch. Paul, E.W. Scheidt, A. Gribanov, Yu. Seropegin, H. Noël, M. Sigrist, and P. Rogl, Phys. Rev. Lett. **92**, 027003 (2004)

[18] Y. Yanase and M. Sigrist, J. Phys. Soc. Jpn. **76**, 043712 (2007)

[19] N. Hayashi, K. Wakabayashi, P.A. Frigeri, M. Sigrist, Physical Review B **73**, 092508 (2006)

[20] E. Bauer, G. Hilscher, H. Michor, C. Paul, E.W. Scheidt, A. Gribanov, Y. Seropegin, H. Noel, M. Sigrist, and P. Rogl, Phys. Rev. Lett. **92**, 027003 (2004)

[21] N. Kimura, K. Ito, K. Saitoh, Y. Umeda, H. Aoki, and T. Terashima, Phys. Rev. Lett. **95**, 247004 (2005)

[22] I. Sugitani, Y. Okuda, H. Shishido, T. Yamada, A. Thamizhavel, E. Yamamoto, T.D. Matsuda, Y. Haga, T. Takeuchi, R. Settai, and Y. Onuki, J. Phys. Soc. Jpn. **75**, 043703 (2006)

[23] K. Togano, P. Badica, Y. Nakamori, S. Orimo, H. Takeya, and K. Hirata, Phys. Rev. Lett. **93**, 247004 (2004)

[24] T. Akazawa, H. Hidaka, T. Fujiwara, T.C. Kobayashi, E. Yamamoto, Y. Haga, R. Settai, and Y. Onuki, J. Phys. Condens. Matter 16, L29 (2004)

[25] J. K. Hulm and R.D. Blaugher, J. Phys. Chem. Solids **19**, 134 (1961)

[26] G. Amano, S. Akutagawa, T. Muranaka, Y. Zenitani, and J. Akimitsu, J. Phys. Soc. Jpn. **73**, 530 (2004)

[27] H. J. Fink, A. C. Thorsen, E. Parker, V. F. Zackay, and L. Toth, Physical Review A **138**, 1170 (1965)

[28] L.E. Toth, W. Jeitschko, and C. M. Yen, Journal of the Less Common Metal **10**, 29-32 (1996)

[29] P. Blaha, K. Schwarz, G. K. H. Madsen, D. Kvasnicka, and J. Luitz, WIEN2K 2001, *An Augmented-Plane-Wave Local Orbitals Program for Calculating Crystal Properties* (Vienna: Karlheinz Schwarz, Technical University of Wien)

[30] J. P. Perdew, K. Burke, and M. Ernzerhof, Phys. Rev. Lett. **77,** 3865 (1996)

[31] N.R. Werthamer, E. Helfand, and P.C. Hohenberg, Phys. Rev. **147**, 295 (1966)

[32] M. Tinkham, *Introduction to Superconductivity*, Second Edition, Page 64

[33] J. P. Carbotte, Rev. Mod. Phys. **62**, 1027 (1990)

[34] Z. Q. Mao, M. M. Rosario, K. D. Nelson, K. Wu, I. G. Deac, P. Schiffer, Y. Liu, T. He, K. A. Regan, and R. J. Cava, Physical Review B **67**, 094502 (2003)

[35] A. Junod, in *Studies of High Temperature Superconductors*, edited by A. Norliker (Nova Science, NewYork; 1996) Vol. 19

Table I. Superconducting and other parameters of $Mo_3Al_2C$

| Parameters | unit | $Mo_3Al_2C$ sample | $Mo_3Al_2C$ [Ref 10,27] | $MgCNi_3$ [Ref 34] |
|---|---|---|---|---|
| $T_c$ | K | 9.2 | 9-10 | 7.63 |
| $\mu_0 H_{c1}(0)$ | T | 0.0047 | 0.0091 @1.2 K | 0.01 |
| $\rho_0$ | $\mu\Omega$ cm | 177.25 | | 120.7 |
| $\mu_0 H_c(0)$ | T | 0.14 | 0.17 @ 1.2 K | 0.19 |
| $\mu_0 H_{c2}(0)$ | T | 18.2 | 15.6 @1.2 K | 14.4 |
| $\xi(0)$ | Å | 42.3 | | 46 |
| $\lambda(0)$ | Å | 3755 | | 2480 |
| $\kappa(0)$ | | 88.56 | | 54 |
| $\gamma(0)$ | mJ/mol $K^2$ | 18.65 | | 30.1 |
| $\Delta C/\gamma T_c$ | | 2.14 | | 2.1 |
| $\mu_0 H^{pauli}$ | T | 16.65 | | 14 |
| $2\Delta_0 / k_B T_c$ | | 4.028 | | 4.4 |
| $\mu_0 (dH_{c2}/dT)_{T=Tc}$ | T/K | -2.39 | | |
| $\Theta_d$ | K | 338.52 | | 284 |

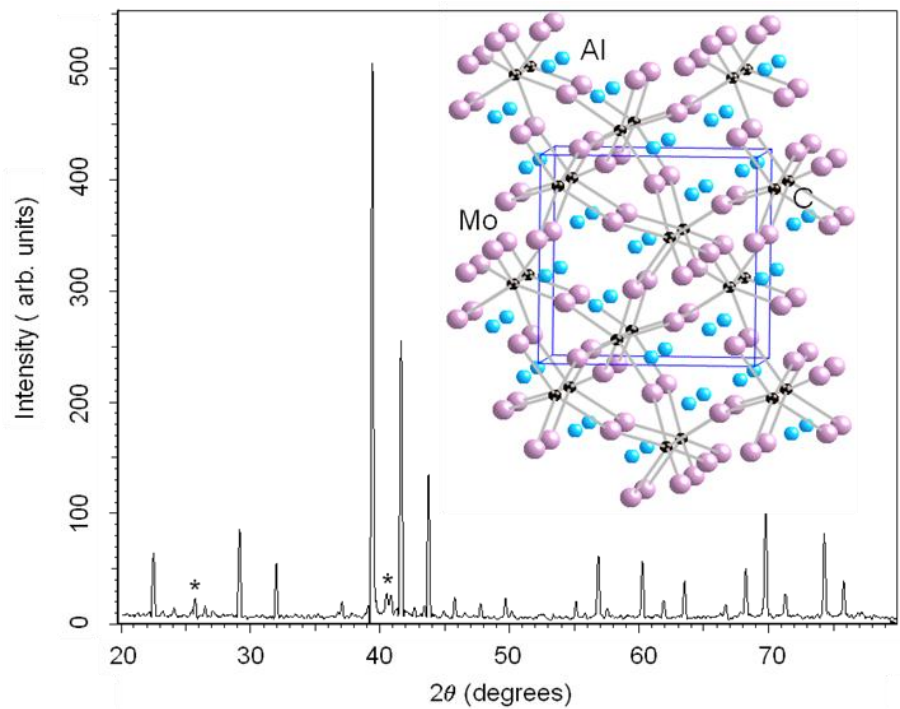

Figure 1. Powder X-ray diffraction pattern and schematic view of the crystal structure of polycrystalline $Mo_3Al_2C$. The *'s in the figure indicate impurity peaks associated with elemental Mo and $Mo_3Al_8$.

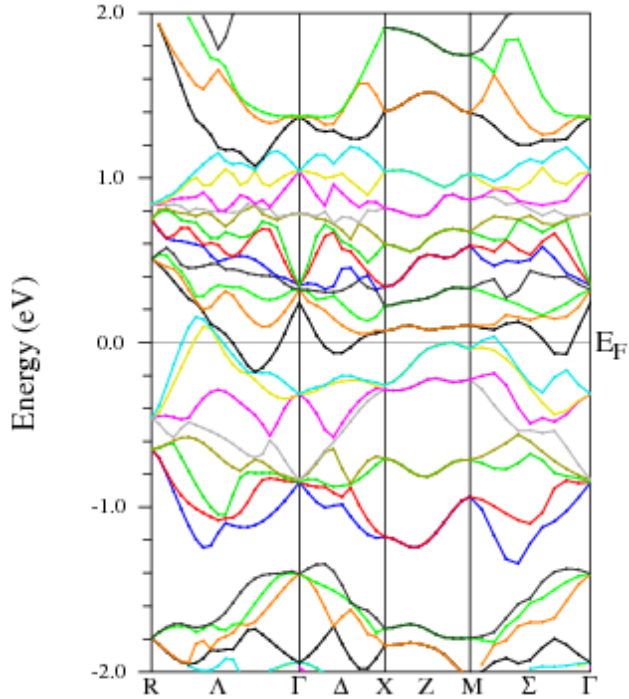

Figure 2. Section of the calculated electronic band structure along high symmetry directions for $Mo_3Al_2C$ in the range of ± 2 around the Fermi energy $E_F$. The band structure is without spin-orbit coupling.

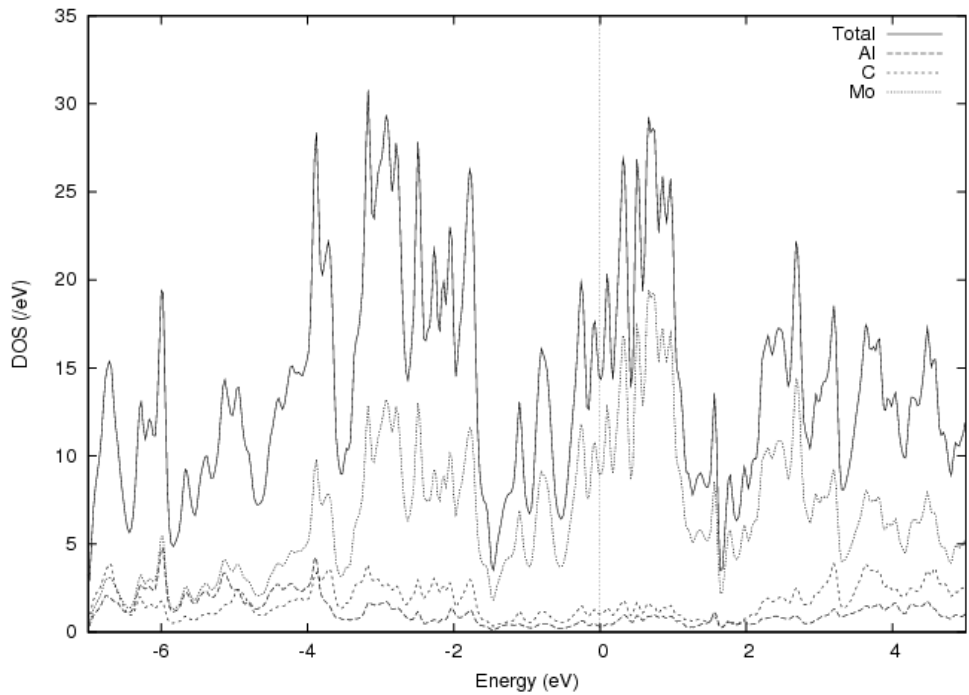

Figure 3. Section of total and atom-projected density of states (DOS; in states $eV^{-1}$) for $Mo_3Al_2C$ in the energy range -7 to +5 around the Fermi energy $E_F$. Energy scale is defined relative to $E_F$. The DOS is shown for scalar relativistic calculation only.

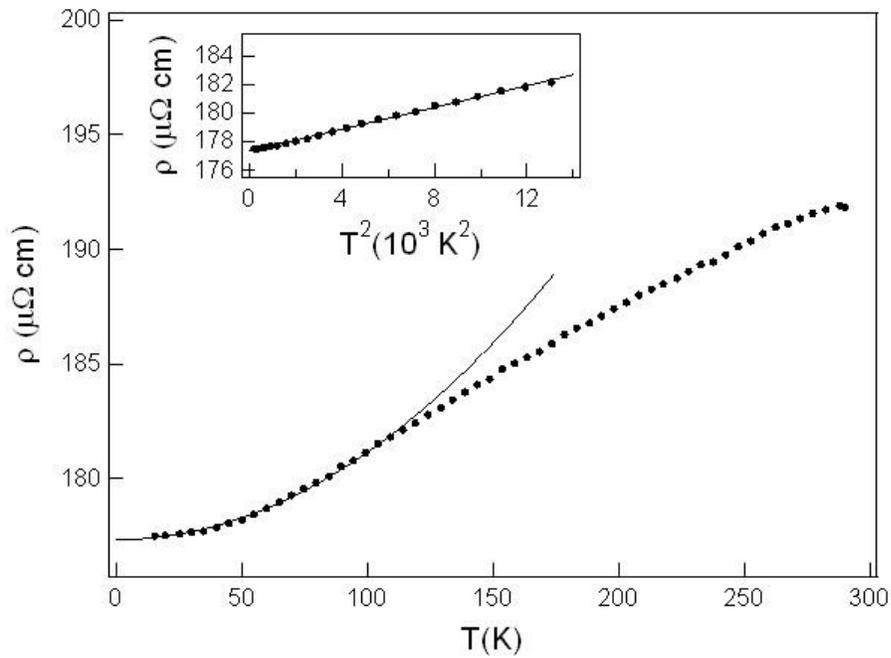

Figure 4. The temperature dependence of the normal state resistivity for a sample of $Mo_3Al_2C$. The solid lines are fits to $\rho(T)$ as indicated in the text.

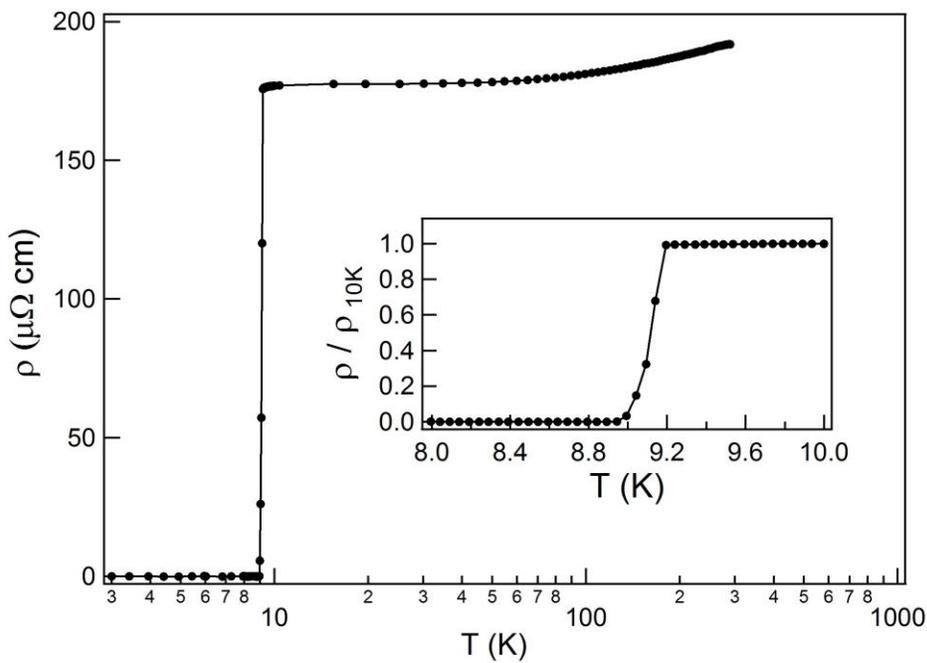

Figure 5. Semi-log plot of the resistivity of polycrystalline $Mo_3Al_2C$ over the range of temperature from 3 to 290 K. The inset shows the superconducting transition at low temperature. The solid lines in both figures are a guide to the eye.

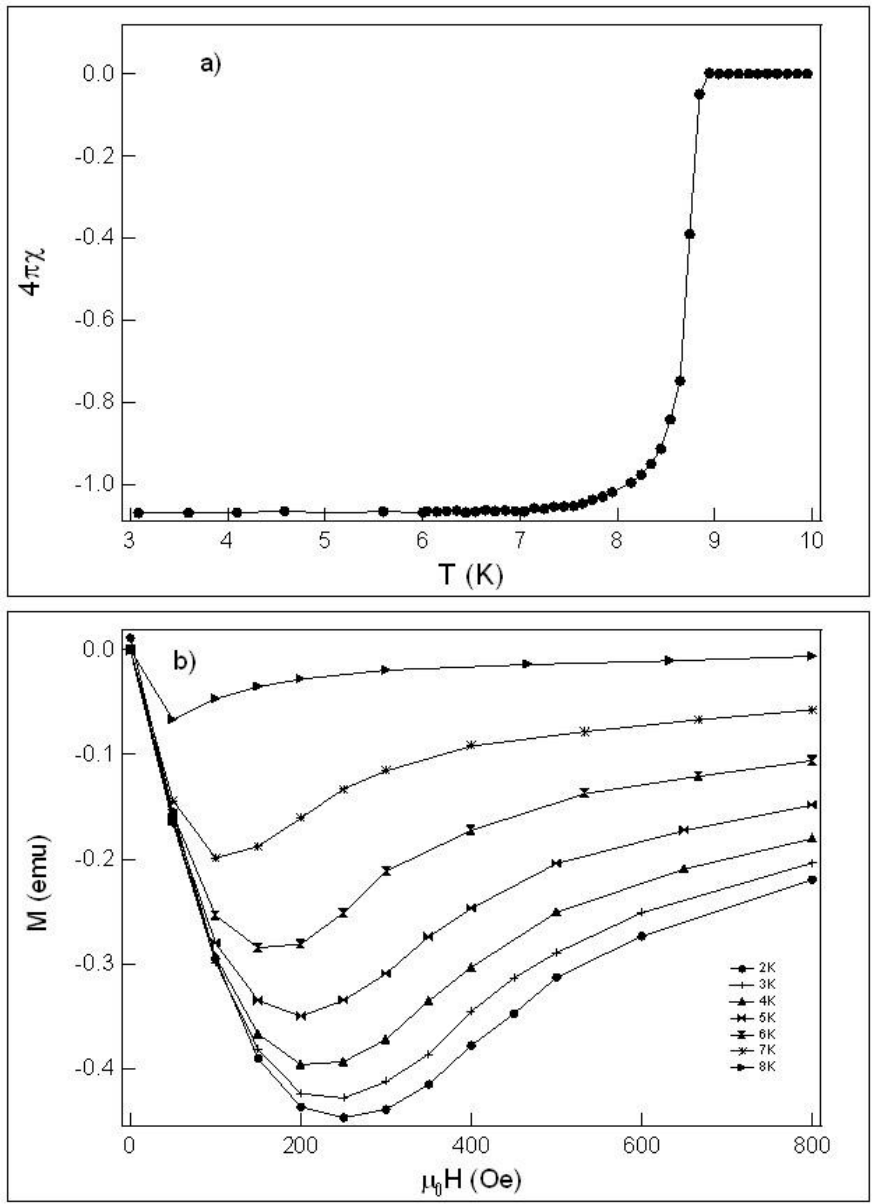

Figure 6. a) ZFC magnetic susceptibility measured at constant magnetic field of 30 Oe. b) Magnetization data with applied magnetic field at various temperatures.

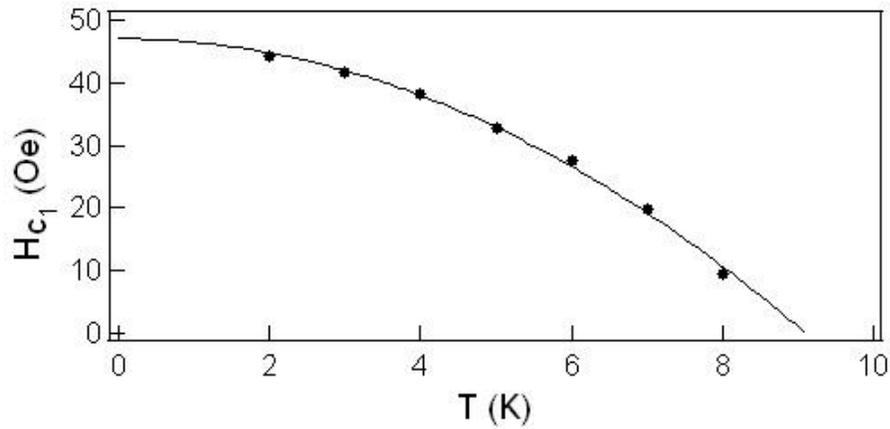

Figure 7. Lower critical field determined by the magnetization measurements at different temperatures. The solid line is the best fit to Equation 2.

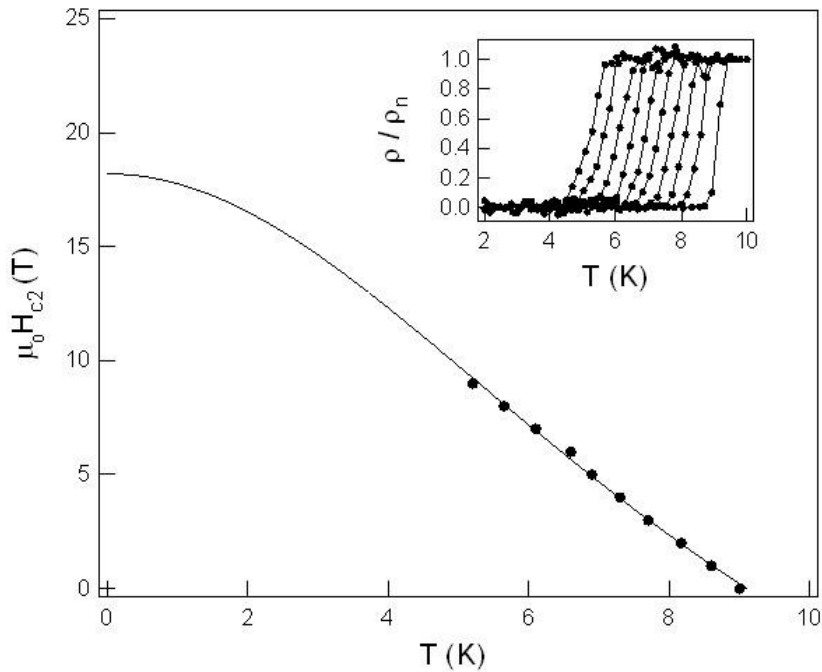

Figure 8. Upper critical field of $Mo_3Al_2C$ as a function of temperature. The solid line is fit to Ginzburg-Landau theory (see text for details). Inset: Characterization of the superconducting transition under magnetic fields 0 T to 9 T from right to left.

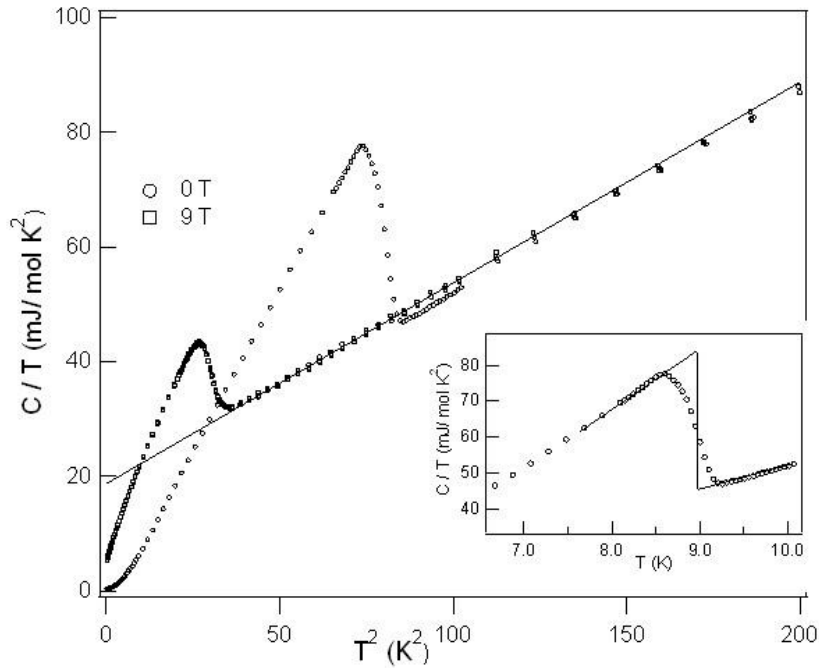

Figure 9. Characterization of $Mo_3Al_2C$ by the measurement of specific heat at 0 and 9 T. Inset: $C/T$ vs. $T$ plot to determine the specific heat jump at the superconducting transition.

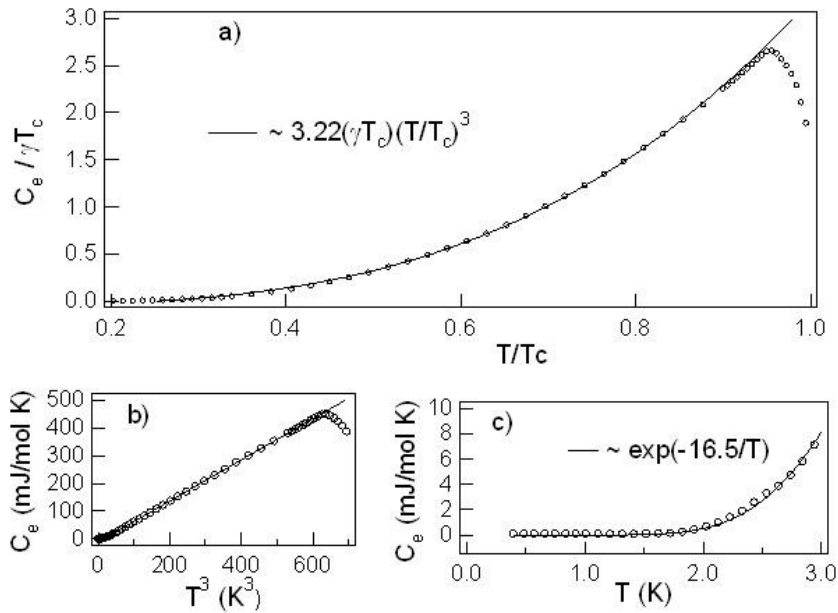

Figure 10. a) Superconducting electronic specific heat capacity ($C_{es}$) vs. reduced temperature ($T/T_c$) of $Mo_3Al_2C$. b) $C_{es}$ vs. $T^3$ plot and c) $C_{es}$ vs. $T$ at low temperature.